\documentclass[a4paper, 12pt]{article}
\usepackage{epsfig}
\usepackage{graphicx}
\usepackage{amsmath}
\begin{document}
\title{Complex Network Structure of Flocks in the Standard Vicsek Model}
\author{Gabriel Baglietto$^{1,2,3}$, Ezequiel V. Albano$^{1,4}$ 
and Juli\'an Candia$^{1,5}$\\
$^1${\small\it Instituto de F\'{\i}sica de L\'{\i}quidos y Sistemas 
Biol\'ogicos (CONICET, UNLP),}\\{\small\it 59 Nro 789, 1900 La Plata, Argentina}\\
$^2${\small\it Facultad de Ingenier\'{i}a (UNLP), La Plata, Argentina,}\\
$^3${\small\it Physics Laboratory, Istituto Superiore di Sanit\`{a},}\\{\small\it Gr. Coll. Roma I, V.le Regina Elena 299,
00161 Roma, Italy}\\
$^4${\small\it Departamento de F\'{\i}sica, 
Facultad de Ciencias Exactas (UNLP),}\\ 
{\small\it La Plata, Argentina}\\
$^5${\small\it Department of Physics, University of Maryland, College Park, MD 20742, USA}\\ 
}
\maketitle

\begin{abstract}
In flocking models, the collective motion of self-driven individuals leads to the formation of complex spatiotemporal patterns. The Standard Vicsek Model (SVM) considers individuals that tend to adopt the direction of movement of their neighbors under the influence of noise. 
By performing an extensive complex network characterization of the structure of SVM flocks, we show that flocks are highly clustered, assortative, and non-hierarchical 
networks with short-tailed degree distributions. Moreover, we also find that the SVM dynamics  
leads to the formation of complex structures with an effective dimension higher than that of the space where the 
actual displacements take place. Furthermore, we show that these structures are capable of sustaining
mean-field-like orientationally ordered states when the displacements are suppressed, thus suggesting a linkage 
between the onset of order and the enhanced dimensionality of SVM flocks. 
\end{abstract}

\section{Introduction}

Nature presents us with an astonishing variety of swarming and flocking phenomena spanning a huge range of lengthscales and 
organismal complexity, from bacterial colonies and migrating cells to insects swarms~\cite{buhl06,szab06,soko07,wu09,zhan09}, from fish schools and shoals to bird flocks and 
mammal herds~\cite{inad02,couz02,giug09,cava10,nagy10}, extending even to crowd dynamics in human collective behavior~\cite{helb00,fari10}. 
Therefore, it is hardly surprising the enormous interest that swarming phenomena has attracted across scientific 
disciplines, involving not just biologists, but also mathematicians, physicists, and engineers. 
Indeed, the modeling of swarming and flocking behavior contributes to the understanding of natural phenomena and 
becomes relevant for many practical
and technological applications as well, e.g. collective robotic motion, design and control of 
artificial microswimmers, microscopic chemical robots (also known as {\it chobots}), 
etc.~\cite{rapp99,wu00,ther02,fede07,gran11} 
(see Ref.~\cite{vics10} for a recent review).

In this broad context, the model early proposed by Vicsek {\it et al}~\cite{vics95}, i.e. 
the so-called Standard Vicsek Model (SVM), has gained large
popularity within the Statistical Physics community, which uses it as an archetypical 
model to study the onset of order upon
the interactive  displacement of self-driven individuals.
The SVM assumes that individuals tend to align their direction of movement when they 
are placed within a certain
interaction range. This rule, which would trivially lead to a fully ordered collective motion, 
is complemented by a second
one that introduces noise in the communications (interactions) among individuals. 
The interplay and competition between these simple rules leads to the observation of a rather complex and interesting nonequilibrium behavior: 
an ordered phase of collective
motion is found for low enough levels of noise, while  a disordered phase is observed at 
high noise levels. Further interest in the
SVM arises from the fact that the nature of the phase transition between those phases could 
be physically described as first- or second-order,
depending on the type of noise considered~\cite{vics95,greg04,alda07,bagl09,bagl12}.

The SVM, which describes a far-from-equilibrium phenomenon, has been compared with the XY model, 
a widely studied critical system that operates under equilibrium conditions~\cite{alda03}. 
The XY model considers 
nearest-neighbor interacting spins that may adopt any possible orientation depending on the 
interplay between temperature and the nearest-neighbor interaction strength~\cite{binn92,chai95}. 
By interpreting the SVM as a model of 
interacting ``spins" that can undergo displacements in the direction of the
spin, the basic difference between both models lies precisely in the SVM's spin displacements. In fact, 
other relevant symmetries for the study
of phase transitions, such as the dimensionality of the space, the nature of the order parameter, 
and the range of the
interactions, are the same in both models. On the other hand, it is well known that, according 
to the Mermin-Wagner Theorem~\cite{merm66,merm67,cass92}, the XY model,
as well as other equilibrium systems sharing the same symmetries, cannot exhibit large-scale 
ordered states in $2$ dimensions.
Therefore, the onset of ordering in the SVM in $2D$ is quite intriguing and has become the subject 
of several studies. Using a variety of approaches such as hydrodynamic equations~\cite{tone95}, long-range links in 
ad-hoc complex network substrates~\cite{alda07}, and off-lattice simulations~\cite{bagl09b}, it has been shown that particle displacements 
in the SVM play the role of effective long-range interactions.   

Within this context, the aim of this paper is to investigate the structural characteristics 
of the complex networks formed by clusters of self-driven individuals during the far-from-equilibrium stationary states of the SVM. 
By means of standard topological measures borrowed from the growing field of complex network research, 
we perform an extensive characterization of the 
structure of SVM flocks and link our findings to intrinsic 
characteristics of the SVM dynamics.  
Furthermore, we also investigate how the complex networks 
formed as a consequence of the particle displacements can still support the
onset of orientational ordering once those displacements are suppressed (i.e. after {\it freezing} the clusters).  
Since, after suppressing the displacements of the individuals, the SVM is essentially analogous to the XY model defined on complex network substrates, we interpret the onset of local ordering in terms of topological features of the frozen clusters that form the ``spin" system's substrates. 

This manuscript is organized as follows: in Section 2, we define the model, 
describe the computer simulation method and how the complex networks are obtained;  
Section 3 is devoted to the presentation and discussion of the results; while our conclusions are stated in Section 4. 
Finally, the Appendix presents the analytic calculation of the clustering coefficient for SVM flocks in the 
large-cluster asymptotic limit.

\section{The Standard Vicsek Model}

The Standard Vicsek Model (SVM) is perhaps the simplest model that captures the essence of collective motion in a 
non-trivial way~\cite{vics95}. 
It consists of a fixed number of interacting particles, $N$, which are moving on a plane. 
The particles move off-lattice with constant and common speed $v_0\equiv |\vec{v}|$.
Each particle interacts locally and tends to adopt the direction of motion of the subsystem of neighboring particles 
(within an interaction circle of radius $R_0$ centered in the considered particle). Since the interaction radius is the 
same for all particles, we define it as the unit of length 
throughout, i.e. $R_0\equiv 1$. 

The updated direction of motion for the $i-$th particle, $\theta_i^{t+1}$, is given by  
\begin{equation}
\theta_i^{t+1}=Arg\left[\sum_{\langle i,j\rangle}e^{i\theta_j^t}\right]+\eta\xi_i^t\ ,
\label{anterm}
\end{equation}
\noindent where $\eta$ is the noise amplitude, the summation is carried over all particles within the 
interaction circle centered at the $i-$th particle, and $\xi_i^t$ is a realization of a $\delta$-correlated white noise uniformly 
distributed in the range between $-\pi$ and $\pi$. 
The noise term can be thought of as due to the error committed by the particle when trying
to adjust its direction of motion to the averaged direction of motion of their neighbors. 
Although several variations to the SVM have later been considered in the literature, such as different noise types, 
models without alignment rule, adhesion 
between neighbor individuals, bipolar particles, etc. (see~\cite{vics10} for a review), in this work we 
focus on the original SVM as formulated by Vicsek {\it et al} in their seminal article~\cite{vics95}.

We implement the model dynamics by adopting the so-called {\it backward update rule}: after the 
position and orientation of all particles are determined at time $t$, we update the position of the particles at time $t+1$ 
according to
\begin{equation}
\vec{x_i}^{t+1}=\vec{x_i}^t+\vec{v_i}^t\ ,
\label{BU}
\end{equation}
\noindent which is then followed by the update of all velocities at time $t+1$ according to Eq.(\ref{anterm}). 
For a detailed discussion on the impact of different updating rules, see Ref.~\cite{bagl09}. 

The SVM exhibits a far-from-equilibrium phase transition between ordered states of motion at 
low noise levels and disordered motion at high noise levels. This order-disorder transition is manifested by the 
natural order parameter of the system, namely the absolute value of the 
normalized mean velocity of the system, given by
\begin{equation}
\varphi = \frac{1}{N v_0}\vert\sum_{i=1}^{N}\vec{v_i}\vert,
\label{orpa}
\end{equation}
\noindent where $\varphi$ is close to zero in the disordered phase and grows up to one in the ordered phase. 
Although the topic remains somewhat controversial, evidence suggests that the phase transition associated with the onset of large-scale ordered flocks is second-order, at least for the type of noise and update rule used in this work. 
For recent discussions on the subject, see e.g. Refs.~\cite{bagl09,nagy07,alda09}.

The Standard Vicsek Model is studied by means of simulations implemented as a cellular automaton, 
where all particles update their states simultaneously in one time step.
The particles move off-lattice in a $2D$ square of side $L=\sqrt{N/\rho}$, where $\rho$ is the particle density. 
We adopted $N=32768$, $v_0=0.1$, and $\rho=0.25$ throughout. For these parameter values (which are standard in the Vicsek model literature), the critical point takes place at the noise amplitude $\eta_c=0.134$~\cite{bagl08}, although we explored a range of other noise values as well. It 
should be also pointed out that scaling relations near the transition region have been reported, which 
therefore link the behavior of 
different model parameters. For instance, noise amplitude and density at criticality are known to scale as $\eta\sim\sqrt{\rho}$~\cite{bagl08}.    
Since we were interested in stationary configurations, we started out our simulations with random initial states and 
disregarded the first $2\times 10^6$ time steps. 
As pointed out in Ref.~\cite{bagl09b}, the order parameter remains unchanged by taking any smaller value for the velocity amplitude $v_0$ 
(which only affects the duration of the transient period, i.e. the time needed to achieve stationary configurations). 

After reaching the stationary regime, we determined the set of connected clusters by means of the 
Hoshen-Kopelman algorithm \cite{hosh76} adapted for the case of off-lattice systems. 
In order to build a set of complex networks that represent the stationary flocks generated by the SVM dynamics, 
we defined that two individuals were linked if the distance between them was within the interaction radius $R_0$. 
Hence, complex networks representing flocks in the stationary regime are non-weighted and undirected. 

We have also investigated the onset of orientation ordering in so-called {\it frozen clusters} (see Sect. 3.4). 
Clusters of individuals were first generated using the full SVM dynamics, as explained above. However, once the 
stationary flocks were obtained, particle displacements (Eq.(\ref{BU})) were suppressed. 
In these frozen clusters, individuals were still allowed to change their orientation following Eq.(\ref{anterm}), 
with a noise amplitude in the range $0<\eta_f<1$, but their locations in space remained fixed.

\section{Results and Discussion}

Earlier studies on the Vicsek Model have shown that, starting with a disordered initial state in which individuals are 
randomly distributed, the dynamic rules lead to the formation of local structures of interacting 
individuals~\cite{vics95,greg04}. 
These structures, which we will call {\it flocks} or {\it clusters} throughout, are not permanent: their shape and size evolve with time, 
with new individuals and sub-flocks merging with them and, conversely, other individuals and sub-flocks separating from them. 
Indeed, the process of merging and dismemberment of sub-flocks can be regarded as an effective long-range interaction, since via this mechanism 
the information of one part of the system may be carried to a different region of space~\cite{alda07,tone95}.  
Although individual clusters do change with time, the statistical properties of the ensemble of clusters are constant once 
we disregard the initial transient regime. Hence, 
at any given time, statistical measurements taken over the flock ensemble are representative of the Vicsek Model's stationary state. 
Besides this statistical perspective, in which flocks are the fundamental building blocks of the flock ensembles that characterize SVM 
stationary states, flocks can be regarded as ``domains" that carry important information on the ordering of the system at the mesoscopic level. 
Bearing in mind these two different perspectives on the role of flocks or clusters as key sub-units within a SVM system, we will 
alternate between the structural analysis of single clusters and the analysis of flocks from the statistical ensemble approach. 

In order to perform an extensive topological characterization of the flocks formed in the stationary phase of the Vicsek Model, 
we will use the complex network approach, which provides us with a conceptual framework and a set of measures that have been applied 
to a huge variety of networked systems from such diverse scientific realms as biology, ecology, sociology, 
physics, computer science, engineering and technology, finance and economics, and others~\cite{newm06,cald07,newm10}.  
By means of this approach, we identify connected clusters or flocks as the basic units of the system and measure 
their structural properties following well-established procedures from the complex network literature. 
As will be shown below, these measurements allow us to characterize the topology of SVM flocks, 
relate them to other networked systems, and understand their ability to sustain ordered states. 

\subsection{Cluster Structure and Size Distribution}

\begin{figure}[t!]
\centerline{{\epsfysize=3.1in\epsfbox{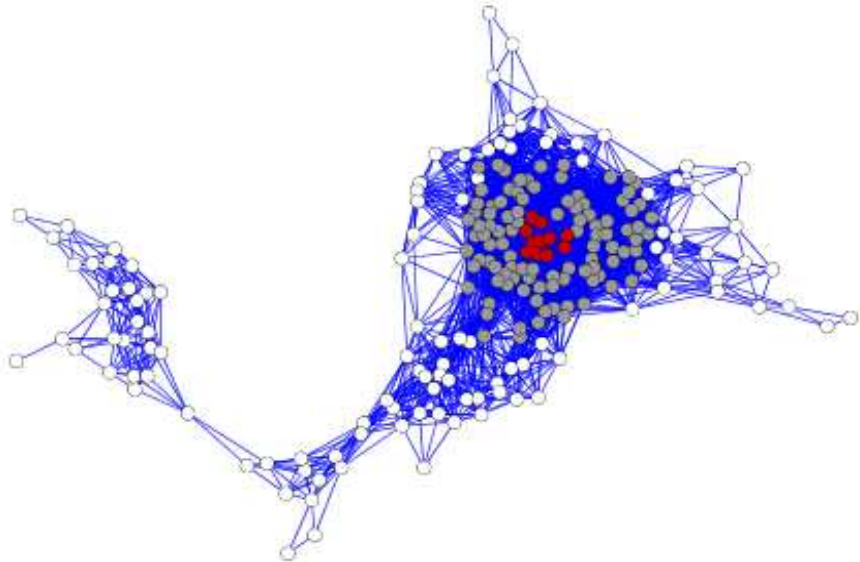}}}
\caption{Complex network structure of a typical SVM flock with 226 nodes, 4147 links, and mean degree $\langle k\rangle=36.7$. 
The length of the links does not represent the actual physical distance between 
neighbor particles. Node colors indicate their degree: white ($k/\langle k\rangle<1$), 
grey ($1\leq k/\langle k\rangle<2$), and red ($2\leq k/\langle k\rangle<3$).
This complex network visualization was created with Cytoscape~\cite{shan03}.}
\label{NetworkMap}
\end{figure}

Figure~\ref{NetworkMap} shows the complex network structure of typical 
flocks in the stationary regime of the Standard Vicsek Model. 
In this complex network representation, the 
length of the links does not correspond to the actual physical distance between 
neighbor particles. 
Notice, however, that the intrinsic modularization of the network structure 
carries significant spatial information. For instance, one can observe 
two distinct modules that correspond to actual sub-flocks  
merging into (or separating from) each other. 
Indeed, flock collision and dismemberment have been identified as 
mechanisms that play a major role in the SVM dynamics~\cite{bagl09}. 

\begin{figure}[t!]
\centerline{{\epsfysize=3.3in\epsfbox{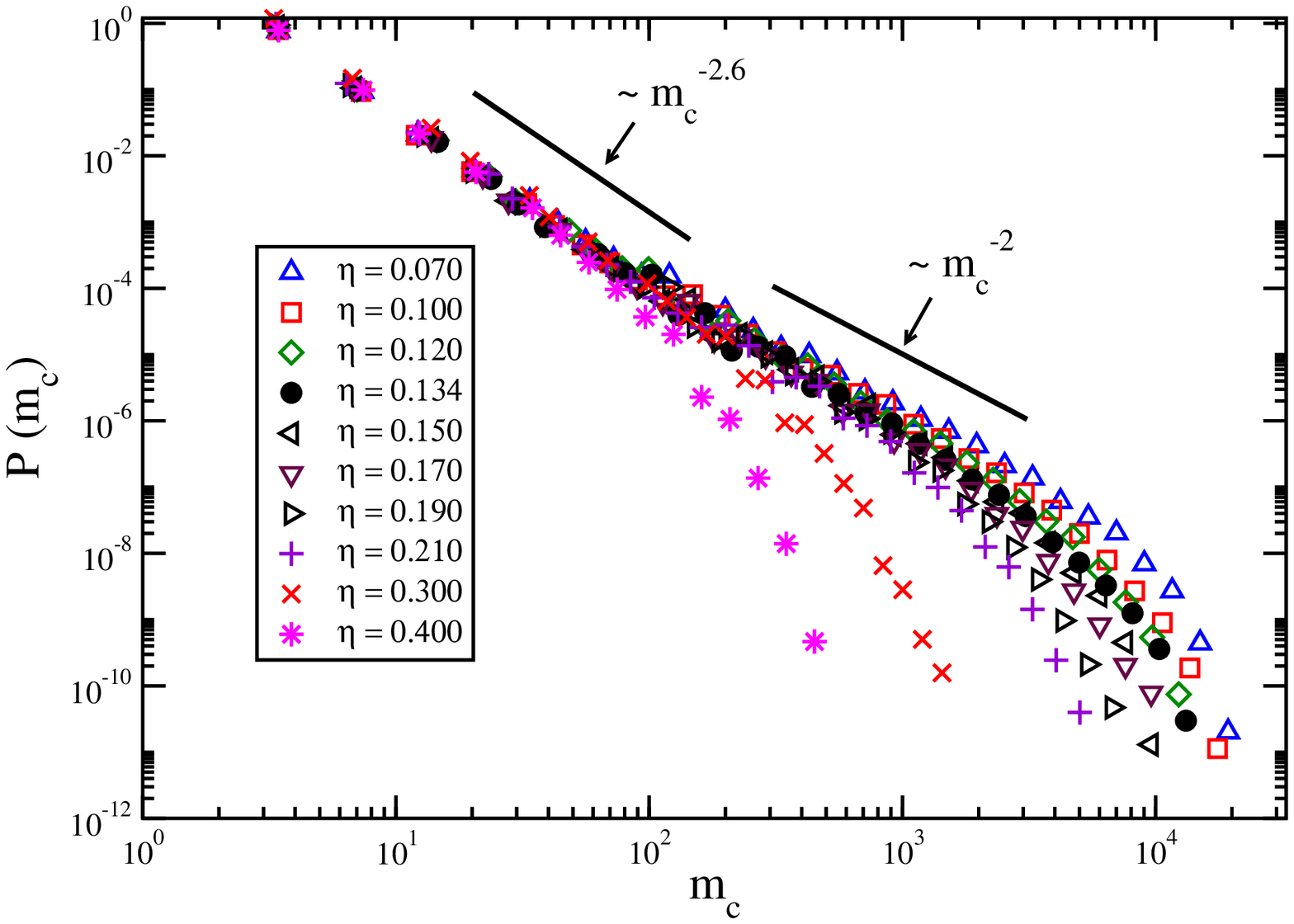}}}
\caption{Cluster mass probability distributions for different levels of noise, 
as indicated. The critical noise is $\eta_c = 0.134$.}
\label{ANClusterDist}
\end{figure}

Nodes are colored according to their degree: 
nodes with fewer connections than average ($k/\langle k\rangle<1$) are 
shown in white, those slightly more connected than average 
($1\leq k/\langle k\rangle<2$) in grey, and the highly connected ones 
($2\leq k/\langle k\rangle$) in red. 
In the flock of Figure~\ref{NetworkMap}, 
$95\%$ of the nodes fall in the first 
two categories, and there are no hubs with $k/\langle k\rangle\geq 3$.  
We observe a central core formed by the highly connected nodes, which is  
wrapped within layers of nodes that are less and less connected 
towards the flock periphery. There are no distinguishable nodes that 
monopolize most of the links (i.e. network hubs), meaning that there are 
no leaders guiding the flock as a whole.   
Moreover, we observe that leaves (i.e. nodes with just one neighbor) 
are very uncommon: only one leaf exists in the graph of Fig.~\ref{NetworkMap}, 
which appears on the far left. 
Notice also the abundance of triangles, which indicates high local clustering. 

In SVM stationary configurations, connected clusters 
(such as the network shown in Fig.~\ref{NetworkMap}) are observed over a very wide range of 
sizes. The probability 
distribution of cluster masses is shown in Figure~\ref{ANClusterDist} for different noise values, as indicated. 
Notice that, here and throughout this paper, we define the mass of a connected cluster, $m_c$, as the number of 
its constituent nodes. 
Over a wide range of noise values both below and above the critical 
point $\eta_c=0.134$, the distributions follow power-laws that cross over to 
exponential decay tails. 
As expected, the power-law span is larger for smaller noise values. 
The exponents that characterize the power-law distributions 
$P(m_c)\sim m_c^{-\beta_c}$ lie in the range $2\leq\beta_c\leq 2.6$. 
For much smaller systems, Huepe and Aldana~\cite{huep04} had found 
$\beta_c\simeq 1.5$ (for $N=500$) and $\beta_c\simeq 1.9$ (for $N=5000$), 
so we conclude that the mass cluster distribution exponents exhibit 
rather strong finite-size effects.  

\begin{figure}[t!]
\centerline{{\epsfysize=3.8in\epsfbox{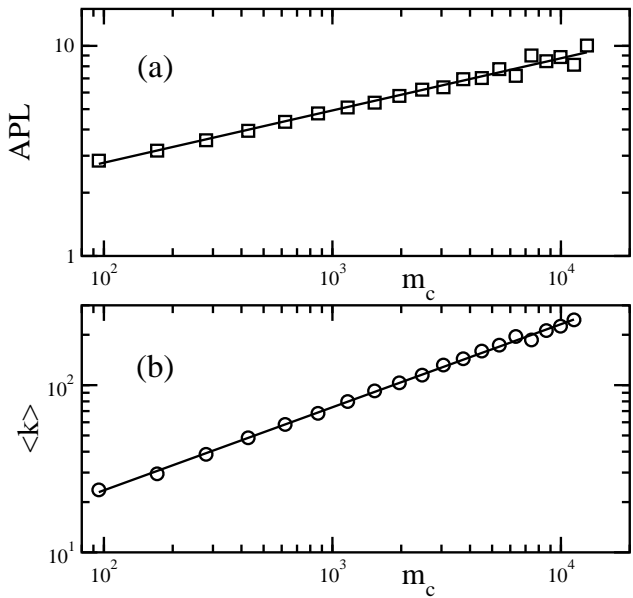}}}
\caption{(a) Log-log plot of the average path length $APL$ as a function of the cluster 
size $m_c$ with $\eta_c = 0.134$. The fit of Eq.(\ref{D_def}) to the data yields $D = 4.0(2)$. 
(b) Log-log plot of the average degree $\langle k\rangle$ as a function of the cluster size $m_c$ 
at the critical noise value $\eta_c = 0.134$. By fitting 
Eq.(\ref{k_alpha}) to the data, we obtain $\alpha = 0.50(1)$.}
\label{APL_AvgDeg}
\end{figure}

In order to gain insight into the structure of the clusters, let 
us first evaluate the average path length, $APL$~\cite{albe02,doro08}. 
According to the standard definition used in the study of complex networks, 
for each pair of nodes $(A,B)$ belonging to the same cluster, the path length $\ell_{AB}$
(also known as chemical distance) is given by the minimum number of links 
that one has to use in order to pass from one node to the other.
Notice that, in networks with undirected links, this distance is the same in both directions. 
By calculating all the pairwise node-to-node path lengths in the cluster and taking the 
average, one obtains the $APL$, which 
consequently is a characteristic length of the cluster.

In Euclidean lattices, the volume of an object is related to its characteristic length by an integer power, 
i.e. the dimension of the object. Based on this observation, as well as on the experience gained in the study of fractal objects, it is customary to define the dimension ($D$) of a complex network according to:
\begin{equation}
APL \propto m_c^{1/D}\ ,
\label{D_def}
\end{equation}                                                      
\noindent where $m_c$ is the complex network size or, in the present context, the 
cluster mass~\cite{doro08}. 
Figure~\ref{APL_AvgDeg}(a) shows a log-log plot of the $APL$ versus $m_c$ for SVM flocks with 
$\eta_c= 0.134$, i.e. clusters corresponding 
to the critical point of the second-order phase transition. The best fit to the data 
yields $D = 4.0 (2)$, which strongly suggests that the effective 
dimension of the clusters is $D = 4$. 

Another quantity of interest for the characterization of complex networks is the average degree distribution 
as a function of the cluster mass~\cite{albe02,doro08}. 
Figure~\ref{APL_AvgDeg}(b) shows a log-log plot of 
$\langle k\rangle$ versus $m_c$, which demonstrates a power-law behavior, i.e.
\begin{equation}
\langle k\rangle \propto m_c^\alpha,
\label{k_alpha}
\end{equation}
\noindent where $\alpha = 0.50(1)$.  

\begin{figure}[t!]
\centerline{{\epsfysize=3.3in\epsfbox{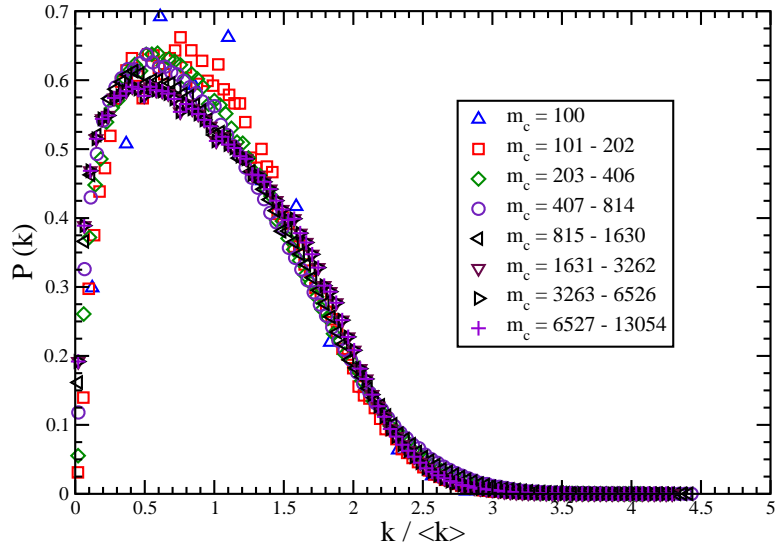}}}
\caption{Degree probability as a function of the normalized degree, 
obtained at the critical noise value $\eta_c = 0.134$ and for different
cluster mass ranges, as indicated.}
\label{DegreeDist}
\end{figure}
                                   
The average degree is only the first moment of a more general property of a network, namely the degree distribution $P(k)$, 
defined as the probability of a vertex to have $k$ links. We evaluated $P(k)$ vs $k$ for different cluster mass ranges, 
as shown in Figure~\ref{DegreeDist} for flocks generated at the critical noise value $\eta_c = 0.134$. 
Notice that, by scaling the horizontal axis by the average degree, all curves collapse, thus indicating 
that clusters of all sizes have a self-similar structure. 
Analogous results (not shown here for the sake of space) are obtained by analyzing clusters in the network ensemble corresponding to the ordered phase (i.e. $\eta<\eta_c$).

The collapsed distribution is short-tailed with most nodes having degrees less than  
$3\langle k\rangle$. Indeed, as noticed above when discussing the structure of 
individual clusters (Fig.~\ref{NetworkMap}), flocks are not guided by leaders, thus the short-tailed 
nature of the degree distribution as opposed to scale-free-like structures 
characterized by the existence of hubs. 
Notice, however, that despite the short-tailed feature, the observed distribution does not match a Poisson distribution. 
In fact, Poisson 
degree distributions are characteristic signatures of classical random graphs, which also show small local 
clustering and a substantial fraction of leaves, while SVM flocks display high local 
clustering and a negligible fraction of leaves. 
We conclude that the short-tailed SVM distribution is a result 
of geographical constraints due to the interaction radius cutoff, which prevents the 
emergence of scale-free topologies.  

\subsection{Effective Dimension of Flocking Clusters}

In the previous Section, we found through Eq.(\ref{D_def}) that SVM clusters 
have an effective dimension $D = 4$. Moreover, the average degree was found to scale 
with the cluster mass as $\langle k\rangle \propto m_c^{1/2}$, Eq.(\ref{k_alpha}).   
Based on these observations, we will now conjecture that these results can be  
rationalized in terms of a projection of a $D = 4$ dimensional object into a $d = 2$ dimensional space. 

\begin{figure}[t!]
\centerline{{\epsfysize=3.7in\epsfbox{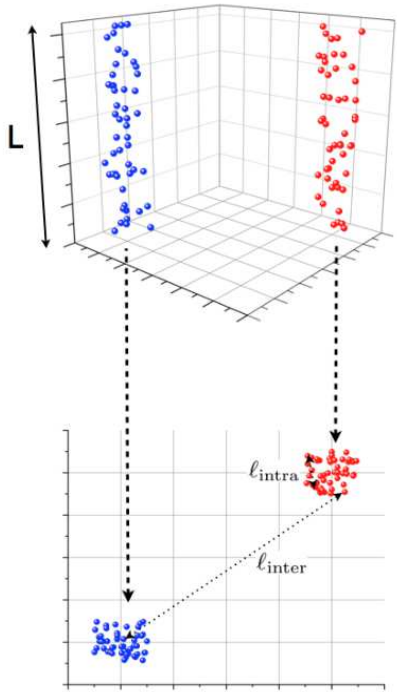}}}
\caption{Schematic representation of the compactification process, which leads to lower-dimensional structures 
with enhanced effective coordination numbers that scale according to Eq.(\ref{k_de_mc}) and average path length 
given by Eq.(\ref{D_def}). Two clusters of nodes are shown in blue and red, as well as the characteristic lengthscales 
$L$, $\ell_{inter}$, and $\ell_{intra}$. See more details in the text. 
}
\label{compactification}
\end{figure}

Let us explore the effective topology of SVM clusters by considering the 
compactification of a hypercubic regular lattice of dimension $D$ and $m_c$ nodes. Hence, we 
begin with a hypercube of side $L=m_c^{1/D}$ and coordination number $k_D$, 
as shown in Figure~\ref{compactification}. If we compactify it 
once, keeping the same number of nodes but projecting them into a hypercube of side $L$ and $D-1$ dimensions, the coordination 
number is increased to $k_{D-1}\approx k_D\times L$. Notice that we can think of this process by assuming 
that each particle (node) in the 
original hypercube is slightly displaced at random off its corresponding lattice site, so that, after the projection, 
we obtain increased local densities without any two particles sharing exactly the same position. Indeed, the compactification process 
generates ``particle lumps" that are responsible for the enhanced effective coordination number.   
After projecting multiple times (from the original $D$ 
dimensions into a $d-$dimensional hypersurface), the coordination number becomes                      
\begin{equation}
k_d\approx k_D L^{D-d},
\label{kd_de_KD}
\end{equation}
\noindent or, by replacing the relation $L=m_c^{1/D}$, 
\begin{equation}
k_d \propto m_c^{\frac{D-d}{D}}.
\label{k_de_mc}
\end{equation} 
On the other hand, we can estimate the average path length of the compactified $d-$dimensional object. 
This object has side $L$ and lumps formed by $n_d=m_c/L^d$ nodes around each one of the $L^d$ lattice sites. 
As shown in Figure~\ref{compactification}, the distance between nodes that belong to the same lump 
is of the order of unity: $\ell_{\rm intra}\sim {\mathcal{O}}(1)$. 
However, node pairs that belong to different lumps have $\ell_{\rm inter}\sim {\mathcal{O}}(L)$ on average. 
Thus, 
\begin{equation} 
\ell\sim
\left\{\begin{array}{cl}{\mathcal{O}}(1) 
\ \ \ \ \text{for\ }n_d(n_d-1)/2\times L^d\ \text{node pairs}\\ {\mathcal{O}}(L)
\ \ \ \ \text{for\ }n_d^2\times L^d(L^d-1)/2\ \text{node pairs}\end{array}\right.\ .
\label{dist}
\end{equation}   
Hence, the leading contribution to the average path length is due to node pairs in different lumps and  
$APL\equiv\langle\ell\rangle\propto L$, which agrees with Eq.(\ref{D_def}) after replacing $L=m_c^{1/D}$.  
In this way, by taking $D = 4$ and $d = 2$, the compactification mechanism leads to node clusters 
that have $APL\propto m_c^{1/4}$ (from Eq.(\ref{D_def})) and $\langle k\rangle\propto m_c^{1/2}$ (from Eq.(\ref{k_de_mc})), 
in agreement with the exponents measured for SVM clusters. 
Let us point out that the
behavior of $L$ as a function of $m_c$ signals the presence of a nontrivial scaling law for the areas of the clusters.
In fact, $L \propto m_c^{1/4}$ is compatible with areas that grow as $m_c^{1/2}$ instead of lineary in $m_c$.

Summing up, in this Section we focused our attention on the average degree and characteristic length of clusters of 
different mass. Our heuristic arguments show that SVM 
clusters can be understood as $4-$dimensional networked objects compactified 
into a $2-$dimensional space. As will be discussed below (see Sect. 3.4), these arguments are also useful in order to 
understand the onset of long-range ordering in frozen clusters. 

\subsection{Clustering Coefficient Analysis of SVM Flocks}

A very important topological measure of a complex network is the clustering 
coefficient, $C$~\cite{albe02,doro08}. 
The clustering coefficient for node $i$ with $k_i$ links is defined as
\begin{equation}
C_i = \frac{2 n_i}{k_i(k_i -1)}\ ,
\label{clusq}
\end{equation}
\noindent where $n_i$ is the number of links between the $k_i$ neighbors of $i$. 
Then, the network's clustering coefficient is calculated as the 
average of $C_i$ taken over all vertices, i.e. $C = \langle C_i\rangle$. 
Empirical results over a wide variety of real networks have shown that $C$ is significantly 
higher for most real networks than for corresponding random networks of similar size~\cite{albe02,watt98,doro02}. 
Furthermore, the clustering coefficient of real networks is to a high degree independent of the number 
of nodes in the network. Interestingly, however, the archetypical complex network models predict a marked 
drop of the clustering coefficient with the network size $N$. For instance, classical random graphs have $C=\langle k\rangle/N$, 
while the Barab\'asi-Albert scale-free model leads to $C\sim N^{-0.75}$~\cite{albe02}. 

\begin{figure}[t!]
\centerline{{\epsfysize=2.8in\epsfbox{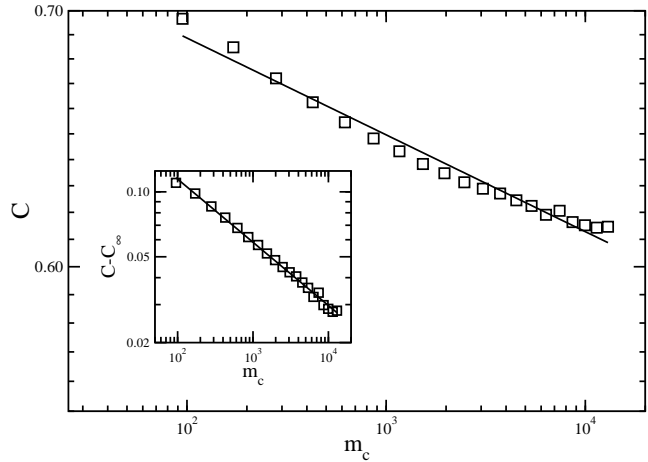}}}
\caption{Log-log plot of the clustering coefficient $C$ as a function of the cluster size $m_c$. 
Inset: after subtracting the asymptotic clustering coefficient $C_\infty$, 
the exponent $\gamma_\infty = 0.30(2)$ is determined.}
\label{CCmassDist}
\end{figure}

Figure~\ref{CCmassDist} shows the dependence of $C$ on $m_c$. 
Notice that flocks of all sizes display a very high degree of clustering, as we anticipated based on the high density of triangles observed in the network structure from Fig.~\ref{NetworkMap}. Also, we observe that the size dependence is very weak: 
fitting the scaling relation $C\propto m_c^{-\gamma}$, we obtain $\gamma=0.025(1)$. 

As shown in the Appendix, the asymptotic clustering coefficient in the limit of an infinitely large cluster, $C_\infty$, can be calculated as a 
function of the density of particles inside the cluster, $\rho_{in}$, according to  
\begin{equation}
C_{\infty} = \frac{[(4\pi -3\sqrt{3})\rho_{in}-8]\pi\rho_{in}}{4(\pi\rho_{in}-1)(\pi\rho_{in}-2)}\ ,
\label{clus_analit}
\end{equation}	
\noindent which is expected to be an 
excellent approximation in the case of large clusters where the surface-to-bulk ratio is negligible.
Since the scaling relation $\langle k\rangle\sim m_c^{0.5}$ implies that $\rho_{in}\to\infty$ 
in the $m_c\to\infty$ limit, Eq.(\ref{clus_analit}) yields 
\begin{equation}
C_{\infty} = 1-\frac{3\sqrt{3}}{4\pi} \simeq 0.587\ .
\label{clusmed}
\end{equation}
\noindent The Inset to Figure~\ref{CCmassDist} shows a log-log plot of $C - C_\infty$ as a function of the cluster mass $m_c$, 
where it is shown that $C - C_\infty$ decays with $m_c$ as a power law with exponent $\gamma_\infty = 0.30(2)$.

\begin{figure}[t!]
\centerline{{\epsfysize=2.8in\epsfbox{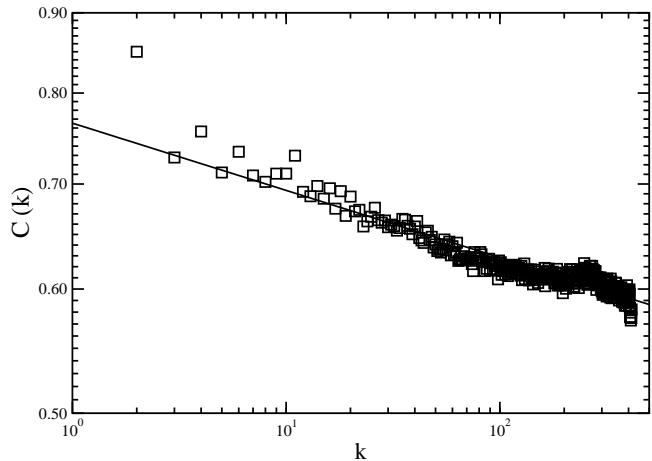}}}
\caption{Log-log plot of the clustering coefficient $C(k)$ as a function of the degree $k$.
Due to their strong geographical constraints, flocks have a nonhierarchical network topology.}
\label{CCdegreeDist}
\end{figure}

In order to determine whether modular organization is responsible for the high clustering coefficients seen in many real networks, 
Ravasz {\it et al}~\cite{rava02,rava03} introduced the scaling law 
\begin{equation}
C(k)\propto k^{-\beta_h}\ ,
\label{hierarchy}
\end{equation}
\noindent where $C(k)$ represents the distribution of the clustering coefficient as a function of the node degree and 
$\beta_h$ is the exponent that measures the hierarchical structure of complex networks. Indeed, it has been observed that 
many real networks are composed of modules that combine into each other in a hierarchical manner. These hierarchical networks are uncovered 
by a scaling behavior of $C(k)$ that follows Eq.(\ref{hierarchy}) with $\beta_h\simeq 1$. 
Figure~\ref{CCdegreeDist} shows a log-log plot of $C(k)$ as a function of the degree. The drop with the degree 
is very mild, namely $\beta_h=0.043(1)$, which is also compatible with a logarithmic decay. This result points to a lack of hierarchical organization 
in the network structure of flocks. We argue that, since links in the network construction process are distance-driven and limited by spatial constraints 
(namely, that particles must lie within an interaction radius $R_0$ in order to be connected), 
the emergence of a hierarchical topology is prevented.  

Another important network characterization is the degree of assortative mixing, i.e. whether high-degree vertices are preferentially attached to other 
high-degree vertices (in which case the network is termed assortative) or whether, on the contrary, high-degree vertices are preferentially attached to 
low-degree ones (in the case of disassortative networks)~\cite{newm02,newm03}. 
Most often, social networks are assortatively mixed, while technological and biological networks 
tend to be disassortative. Network models such as classical random graphs and Barab\'asi-Albert scale-free networks are neither assortative nor disassortative. 

\begin{figure}[t!]
\centerline{{\epsfysize=2.8in\epsfbox{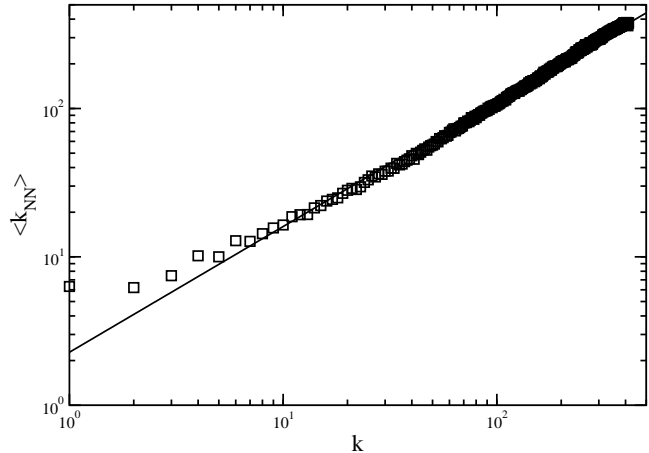}}}
\caption{Log-log distribution of the average degree $\langle k_{NN}\rangle$ of nearest-neighbors of nodes of degree $k$, which reveals 
that SVM flocks have a highly assortative network topology.}
\label{Assortativity}
\end{figure}

One way to determine the degree of assortative mixing is by considering the average degree $\langle k_{NN}\rangle$ calculated among the nearest-neighbors of 
a node of degree $k$. Figure~\ref{Assortativity} shows a log-log plot of the $\langle k_{NN}\rangle$ vs $k$ distribution, which  
reveals a very high degree of assortative mixing. Indeed, by fitting the data to a power-law of the form 
$\langle k_{NN}\rangle\propto k^{\gamma_a}$, the assortativity exponent turns out to be $\gamma_a=0.848(4)$. 
Alternatively, one can measure the degree of assortativity as the Pearson correlation coefficient of the degrees at either ends of an edge. This measure, 
originally introduced by Newman~\cite{newm02}, is obtained from the expression
\begin{equation}
r={{1}\over{\sigma_q^2}}\sum_{ij}ij\left(e_{ij}-q_iq_j\right)\ ,
\label{assort}
\end{equation}
where $i,j$ are the degrees of the vertices at the ends of a given edge and the summation is carried over all edges in the network. Instead of using a node's 
degree $k_i$, here we are interested in the node's {\it remaining degree} $q_i=k_i-1$ that excludes the edge between the two nodes being considered. Moreover, $e_{ij}$ 
is the joint probability distribution of the remaining degrees of the two vertices at either end of a randomly chosen edge~\cite{call01}, and 
$\sigma_q^2=\sum_kk^2q_k-\left[\sum_kkq_k\right]^2$ is the variance of the $q_k$ distribution. 
The definition of $r$ through Eq.(\ref{assort}) lies in the range $-1\leq r\leq 1$, with assortative networks having $r>0$ and disassortative ones having $r<0$. 
For instance, several scientific collaboration networks show assortative mixing in the range $0.12\leq r\leq 0.36$, while the network 
of connections between autonomous systems on the Internet has $r=-0.19$ and the food web from undirected trophic relations in Little Rock Lake, Wisconsin 
has $r=-0.28$~\cite{newm02}.
The Pearson correlation coefficient measured among large flocks turns out to be $r_A=0.82(6)$, i.e. 
SVM flocks have very high assortative mixing. 

As mentioned above, it is well known that high local clustering and high assortativity are distinct hallmarks of social networks. 
Moreover, the imitation mechanism between neighboring interacting particles introduced by the SVM dynamics resembles 
well-studied ``ferromagnetic"-like interactions that play a key role in the occurrence of 
social cooperative phenomena~\cite{bord01,cand07}. Hence, these observed structural properties of SVM flocks can be interpreted as arising from the social nature that underlies the behavior of individuals according to the SVM 
dynamics.      
 
\subsection{Onset of Orientation Ordering in Frozen Clusters}
One of the most intriguing features of the SVM is the 
onset of long-range ordering and the existence of an order-disorder phase transition in $d = 2$ dimensions.
In order to explore this phenomenon, here we analyze whether the topology of frozen clusters, 
once particle displacements and cluster rearrangements are suppressed, is capable by itself of supporting the 
existence of an orientationally ordered phase. For this purpose, we first generate configurations of clusters by applying the 
full SVM dynamics. Once the non-equilibrium stationary state is reached, we identify the clusters 
and ``freeze" them, i.e. we disallow any further displacements of the individuals. From that point on, 
the orientation of the particles is allowed to evolve according to the usual rule (Eq.(\ref{anterm})), 
but subsequent displacements (Eq.(\ref{BU})) do not occur. We will refer to this stage as ``restricted SVM dynamics".  

\begin{figure}[t!]
\centerline{{\epsfysize=3.3in\epsfbox{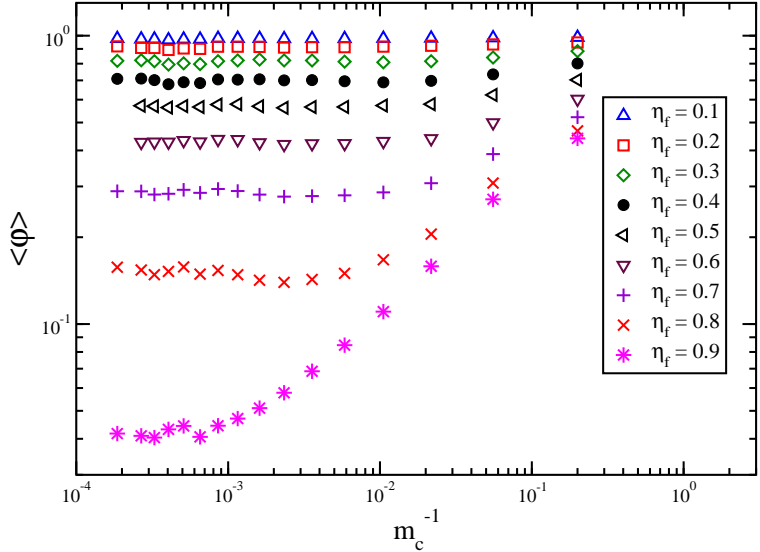}}}
\caption{Log-log plot of the order parameter as a function of the inverse cluster mass for frozen clusters with different noise levels, 
as indicated.}
\label{AN_OPFrozen}
\end{figure}

In this Section, we focus on single flocks, regarding them as domains that carry important information on 
the ordering of the system at the mesoscopic level. With this aim, we explore the relation between local topology and the 
ability for the restricted SVM dynamics to sustain local ordering. 
Clearly, with the full SVM dynamics, the onset of ordering within individual flocks is
required in order to have ordered system-wide macroscopic states. However, the full SVM dynamics has an entanglement between particle 
displacements (``spin" translations) and XY-type interactions (``spin" rotations). 
By resorting to ``frozen clusters", we disentangle these two major components. 
Notice that, although ordering in individual flocks is a necessary condition to have macroscopic system-wide ordering in the full SVM dynamics, 
the sufficient conditions that guarantee macroscopic order are not understood yet. We know that flocks merge, collide, and 
dismember, and it is by virtue of these transport mechanisms that the ordering information within one flock is carried across the system, thus resulting in effective long-range interactions~\cite{alda07,tone95}. 
As shown by Toner et al.~\cite{tone95,tu98}, the spontaneous symmetry breaking of the velocity field leads to 
``Goldstone mode" fluctuations. In equilibrium systems, such fluctuations are strong enough to destroy the long-range order in 2 dimensions. However, the nonequilibrium effect of the nonlinear terms (which violate Galilean invariance, as 
expected due to the existence of a preferential reference frame) stabilize long-range ordering in the continuum model of flocking. Notice also that we are concerned with stationary states (see Section 2 for simulation details), hence ensemble averages are independent of time. While individual flocks (such as the one shown in Figure~\ref{NetworkMap}) change over time, flock ensemble properties (such as the cluster mass distributions shown in Figure~\ref{ANClusterDist}) are stationary.    
However, the detailed 
mechanisms leading to the emergence of global 
order from locally ordered clusters are not well understood yet and remain an open question that lies beyond the scope of this work.  

\begin{figure}[t!]
\centerline{{\epsfysize=3.3in\epsfbox{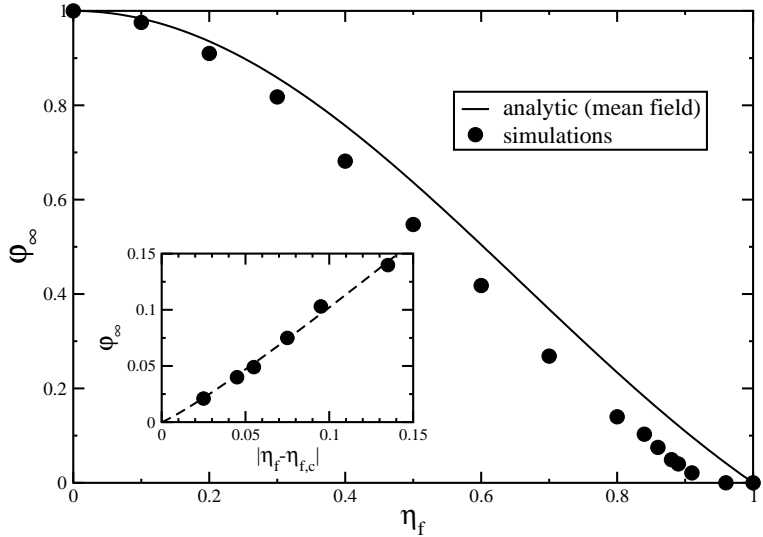}}}
\caption{Plot of the asymptotic values of the order parameter versus the noise amplitude. 
The solid line shows the mean field (i.e. fully-connected graph) results~\cite{doss09}. Inset: Simulation results (symbols) and fit to the data 
(dashed line), from which $\beta=1.03(5)$ and $\eta_{f,c}=0.935(5)$ are obtained.}
\label{asymptotic}
\end{figure}

Figure~\ref{AN_OPFrozen} shows the dependence of the order parameter $\varphi$ on the inverse cluster mass $m_c^{-1}$, as obtained for frozen clusters. 
The clusters were first generated using the full SVM dynamics with critical noise $\eta_c$. After freezing them, 
the restricted SVM dynamics was applied using different noise values in the $0 < \eta_f < 1$ range, as indicated. 
For each $\eta_f$, the corresponding order parameter plot exhibits a plateau in the $m_c^{-1}\ll 1$ region, thus indicating that 
a finite order parameter $\varphi_\infty>0$ persists in the thermodynamic limit. Indeed,  
the order parameter in the large-cluster limit remains positive even for very large noise amplitudes, e.g. $\varphi_\infty\simeq 0.04$ for $\eta_f=0.9$. 

\begin{figure}[t!]
\centerline{{\epsfysize=3.3in\epsfbox{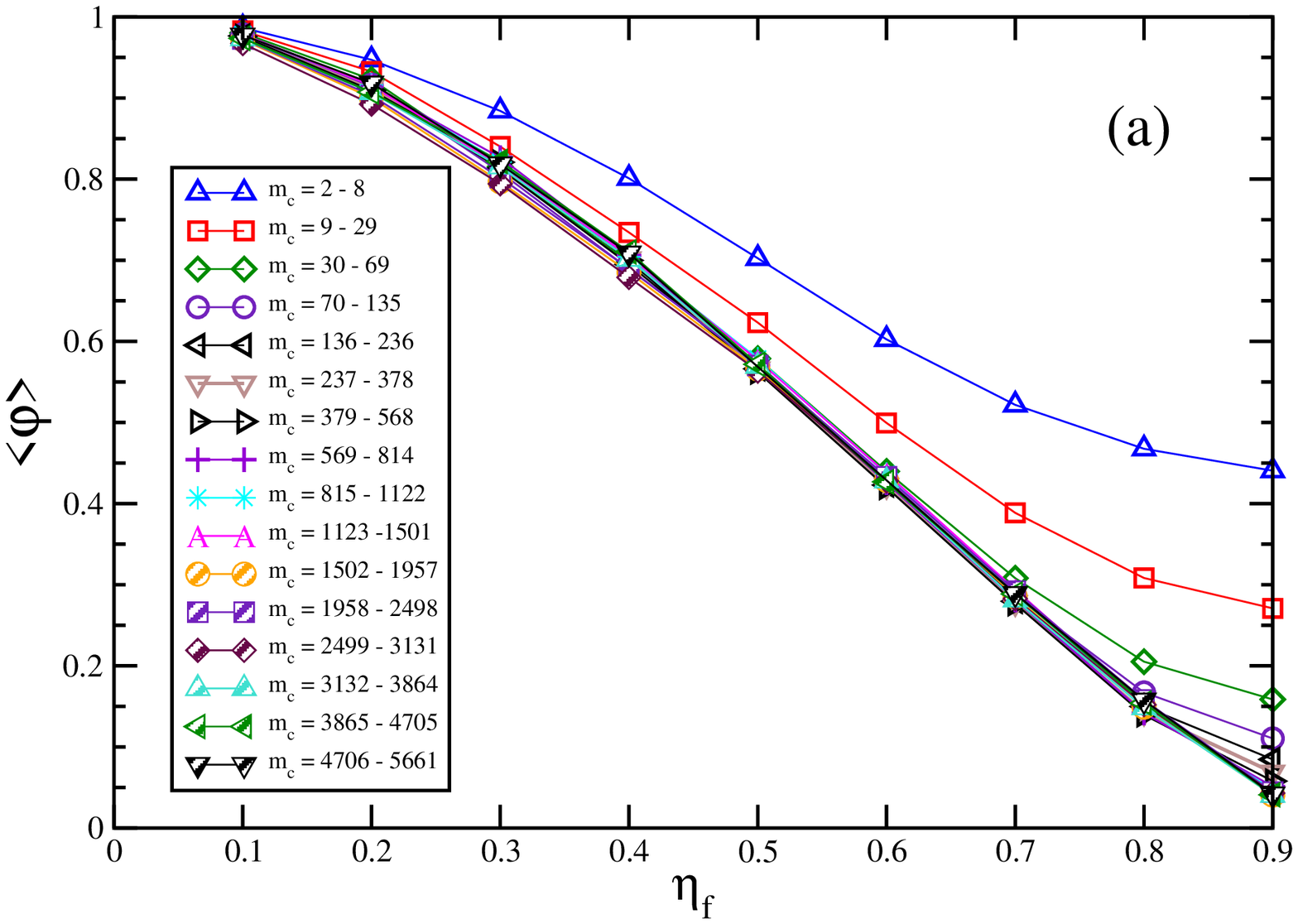}}}
\centerline{{\epsfysize=3.3in\epsfbox{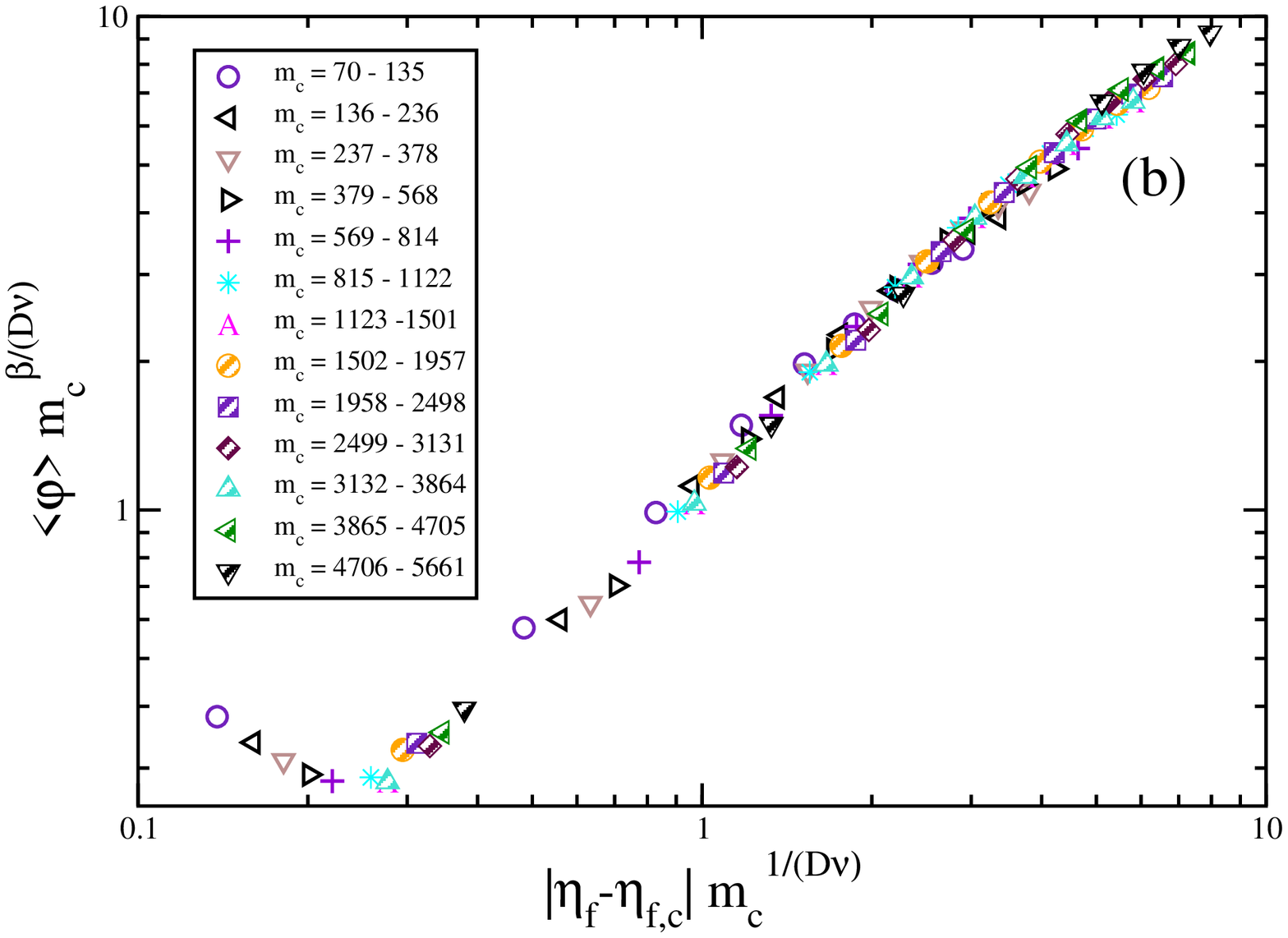}}}
\caption{(a) Order parameter as a function of the noise amplitude $\eta_f$ for SVM frozen clusters grouped 
according to cluster size, as indicated. (b) Data collapse that shows universal behavior according to standard finite-size scaling laws 
of critical systems, using $\eta_{f,c}=0.935$, $\beta=1$, $\nu=1$, and $D=4$.}
\label{FSS}
\end{figure}

Let us now present additional evidence of the orientation ordering sustained by SVM frozen clusters. 
The symbols in Figure~\ref{asymptotic} show the order parameter extrapolations to the $m_c^{-1}\to 0$ limit as a function of the noise amplitude $\eta_f$. For the sake of comparison, this figure also shows exact results from the mean field (i.e. fully-connected graph) 
solution obtained for an infinite density of individuals, namely $\varphi_{MF}= \sin(\pi\eta_f)/\pi\eta_f$~\cite{doss09}, which closely follows the trend of our computer simulation results. The Inset to Figure~\ref{asymptotic} shows a fit to the scaling relation $\varphi_\infty\sim |\eta_f-\eta_{f,c}|^\beta$ (dashed line) in the neighborhood of the critical frozen noise amplitude $\eta_{f,c}$. By fitting our simulation data to this relation, we find that $\beta=1.03(5)$ and $\eta_{f,c}=0.935(5)$. The leading term in the expansion of the mean-field solution around $(\eta_{f,c})_{MF}=1$ leads to $\beta=1$, in excellent agreement with the simulation results.   

Figure~\ref{FSS}(a) shows the order parameter as a function of the noise amplitude $\eta_f$ for SVM frozen clusters grouped 
according to cluster size, as indicated. Except for the very small clusters, we observe that the ordering behavior falls 
into a universal curve that is essentially size-independent. Therefore, the orientation ordering signaled by $\varphi>0$ 
is expected to hold for all noise amplitudes below the critical value $\eta_{f,c}=0.935$ in the thermodynamic limit ($m_c\to\infty$). 
In order to complete this picture, we can perform a finite-size scaling data collapse. From standard finite-size scaling theory, 
we know that $\langle\varphi\rangle L^{\beta/\nu}\sim |\eta_f-\eta_{f,c}|L^{1/\nu}$, where $L$ is a characteristic linear scale of the system and $\nu$ 
is the exponent that characterizes the divergence of the correlation length at criticality~\cite{priv90}. 
Moreover, by combining Josephson, Rushbrooke and Fisher's critical exponent relations (see, e.g., Ref.~\cite{chai95}), we obtain the relation 
$\nu=2\beta/(D-2)$. By replacing 
$\beta=1$ (obtained from expanding the mean-field solution to leading term, which is in agreement with the fit to our data shown in the Inset to Figure~\ref{asymptotic}) and $D=4$ (the effective dimension of 
SVM clusters, as discussed in Section 3.2), we obtain $\nu=1$. Furthermore, we can substitute $L\sim m_c^{1/D}$ in the 
finite-size scaling relation.  
Thus, the critical behavior of the orientation of particles under 
the restricted SVM dynamics should manifest itself as the collapse of data from frozen clusters of different 
sizes when plotted as $\langle\varphi\rangle m_c^{\beta/(D\nu)}\sim |\eta_f-\eta_{f,c}|m_c^{1/(D\nu)}$ with 
$\eta_{f,c}=0.935$, $\beta=1$, $\nu=1$, and $D=4$.
Figure~\ref{FSS}(b) shows the data collapse that confirms the finite-size scaling behavior of the system.      
Based on the excellent agreement between our results and the expected behavior from finite-size scaling theory, we argue that the observed 
mean-field-like behavior is related to the fact that $D = 4$ is the upper-critical dimension of the XY model, which essentially has 
the same symmetries as the SVM defined on frozen clusters.  

\begin{table}[ct!]
\begin{center}
\begin{tabular}{||c||c||} \hline \hline
\multicolumn{1}{||c||}{Observable}&\multicolumn{1}{c||}{Result}\\ \hline \hline
 Cluster size distribution: $P\propto m_c^{-\beta_c}$ & $\beta_c=2-2.6$ \\ \hline
 Average path length: $APL\propto m_c^{1/D}$ & $D=4.0(2)$ \\ \hline
 Average degree distribution: $\langle k\rangle\propto m_c^\alpha$ & $\alpha=0.50(1)$ \\ \hline
 Degree distribution: $P(k)$ vs $k$ & Short-tailed \\ \hline
 Clustering coefficient distribution: $C\propto m_c^{-\gamma}$ & $\gamma=0.025(1)$ \\ \hline
 Asymptotic clustering coefficient: $C_\infty$ & $C_\infty=0.587$ \\ \hline
 Reduced clustering coefficient distribution: & $\gamma_\infty=0.30(2)$ \\
 $C-C_\infty\propto m_c^{-\gamma_\infty}$ & \\ \hline
 Hierarchical modularity: $C(k)\propto k^{-\beta_h}$ & $\beta_h=0.043(1)$ \\ \hline
 Assortative mixing: $\langle k_{NN}\rangle\propto k^{-\gamma_a}$ & $\gamma_a=0.848(4)$ \\ \hline
 Assortative mixing & $r=0.82(6)$ \\ 
 (Pearson correlation coefficient) & \\ \hline
 Frozen clusters: $\lim_{m_c\to\infty}\varphi$ & Mean Field \\ \hline \hline
\end{tabular}
\caption{Summary of results.}
\label{summary}
\end{center}
\end{table}

\section{Conclusions}

In this work, we presented a detailed study of the structural properties of Standard Vicsek Model (SVM) flocks from a complex network perspective. 
The main results are summarized in Table~\ref{summary}. 
The complex network structure of SVM flocks is characterized by a short-tailed degree distribution, very high clustering, 
very high assortative mixing, and nonhierarchical topology. Qualitatively, we can explain these common features as due to 
the intrinsic distance-driven, ``ferromagnetic" nature of the Vicsek model. On the one hand, the interaction radius imposes a cutoff in the 
range of the particle interactions, which is reflected in the nonhierarchical topology of SVM clusters and the short-tailed degree distribution. 
On the other hand, the strong tendency 
among neighbor particles to align with each other, akin to multiple-state ferromagnets such as the XY model and resembling 
typical interaction mechanisms of social networks, leads to very high local clustering 
and assortative mixing. Based on these observations, we can characterize SVM flocks 
as {\it geographically-constrained ``social" networks}.

Furthermore, the average path length dependence on cluster size shows the formation of complex 
structures with an effective dimension higher than that of the space where the actual displacements take place. 
These observations are consistent with assuming SVM clusters as $4-$dimensional networked objects compactified 
into a $2-$dimensional space. Further support to these conclusions comes from our investigation on the onset of ordering in frozen clusters (i.e. when the particle displacements are suppressed).
Indeed, we observe that frozen clusters are capable of sustaining mean-field-like orientationally ordered states (analogously to the XY model in $4D$). This behavior is in sharp contrast with that of 
equilibrium systems in $2D$ space with short-range interactions and $O(2)$ symmetry defined on translationally-invariant substrates, which are indeed prevented from displaying ordered phases due to the Mermin-Wagner Theorem~\cite{merm66,merm67,cass92}.  

Most of the quantitative results in this work were obtained in the critical region of the SVM. A full quantitative analysis 
of the behavior of Vicsek flocks within the ordered phase and its dependence upon the noise amplitude would require a great 
computational effort that lies beyond the scope of the present paper, but remains a promising open question that would certainly deserve 
attention in further work.  

\section*{Acknowledgments}
We are very grateful to F. V\'azquez for fruitful
discussions. This work was financially supported by  CONICET, UNLP and  ANPCyT (Argentina).

\appendix
\section*{Appendix}
\setcounter{section}{1}

Here we derive Eq.(\ref{clus_analit}) for the mean clustering coefficient of a vertex 
in the bulk of a large cluster. Large flocks generally consist of a core that contains most of 
the particles and links between them distributed in a highly uniform fashion. Indeed, the near-uniform 
distribution of nodes and links within flock cores is not only spatial but it also manifests itself in the network's structure, as 
for instance shown by the short-tailed degree distributions in Fig.~\ref{DegreeDist}. Based on these 
observations and for the sake of simplicity, 
we assume that particles in the bulk are distributed homogeneously with a constant density $\rho_{in}$. 

Let us recall that the clustering coefficient for node $i$ with $k_i$ links is defined as $C_i = 2 n_i/(k_i(k_i -1))$, 
where $n_i$ is the number of links between the $k_i$ neighbors of $i$.
Since particles are distributed uniformly within the interaction radius $R_0=1$, it follows straightforwardly 
that $k_i=\pi\rho_{in}-1$.
In order to evaluate $n_i$, let us focus our attention on one of the neighbor nodes of $i$, which we call node $j$. 
The number of nodes that are neighbors of $i$ and $j$ simultaneously is, on average, given by 
\begin{equation}
n_{ij}=\rho_{in}A_{ij}-2\ ,
\label{nij}
\end{equation}
\noindent where $A_{ij}$ is the area of the intersection between the interaction circles centered at $i$ and $j$.
$A_{ij}$, which depends only on the distance $r$ between $i$ and $j$, can be expressed as
\begin{equation}
A_{ij}(r)= 2\int_{-\sqrt{1-\frac{r^{2}}{4}}}^{\sqrt{1-\frac{r^{2}}{4}}}(\sqrt{1-x^{2}}-\frac{r}{2}) \: dx\ .
\label{Aij}
\end{equation}
Therefore, the number of links between the $k_i$ neighbors of $i$ is obtained by replacing Eq.(\ref{Aij}) in Eq.(\ref{nij}) and 
integrating $\rho_{in}n_{ij}/2$ over the unit circle (notice that we divide by 2 because we are dealing with undirected links, so we 
must count each pair of neighbor nodes just once), i.e. 
\begin{equation}
n_i=\pi\rho_{in}\int_0^1 rn_{ij}(r) \:dr \ .
\label{nijr}
\end{equation}
Solving the integrals and replacing in the definition of the clustering coefficient, we finally arrive at 
\begin{equation}
C_\infty = \frac{[(4\pi -3\sqrt{3})\rho_{in}-8]\pi\rho_{in}}{4(\pi\rho_{in}-1)(\pi\rho_{in}-2)}\ ,
\label{clus_analit_app}
\end{equation}
\noindent which provides an analytic solution for the mean clustering coefficient of particles in the bulk of a large cluster.

\end{document}